\documentclass[apj]{emulateapj}
\usepackage{booktabs}
\usepackage{graphicx}% Include figure files
\usepackage{dcolumn}% Align table columns on decimal point
\usepackage{bm}% bold mathp
\usepackage{amsmath}

\usepackage{color}

\def\nar{\ref@jnl{New A Rev.}}          % New Astronomy Review

\def\msun{M_\odot}

\begin{document}

\title{GRB beaming and gravitational-wave observations}
\shorttitle{GRB beaming and GW detection}
\shortauthors{Chen \& Holz}

\author{Hsin-Yu Chen\altaffilmark{1} and Daniel E. Holz\altaffilmark{2}}
\affiliation{$^1$Department of Astronomy and Astrophysics, University of
  Chicago, Chicago, IL 60637\\
$^2$Enrico Fermi Institute, Department of Physics, and Kavli Institute
for Cosmological Physics\\University of Chicago, Chicago, IL 60637}

%\date{\today}% It is always \today, today,
             %  but any date may be explicitly specified

\begin{abstract}
Using the observed rate of short-duration gamma-ray bursts (GRBs) it is possible
to make predictions for the detectable rate of compact binary coalescences in
gravitational-wave detectors. These estimates rely crucially on the growing
consensus that short gamma-ray bursts are associated with the merger of two
neutron stars or a neutron star and a black hole, but otherwise make no
assumptions beyond the observed rate of short GRBs. In particular, our results
do not assume coincident gravitational wave and electromagnetic observations. We show that the
non-detection of mergers in the existing LIGO/Virgo data constrains the
progenitor masses and beaming angles of gamma-ray bursts (e.g.,
$\theta_j>4^\circ$ for $M_{\rm total}\ge20\msun$, for uniform component mass), although these limits
are fully consistent with existing expectations.
We make predictions for the
rate of events in future networks of gravitational-wave observatories, finding
that the first detection of a NS--NS binary coalescence associated with the
progenitors 
of short GRBs is likely to happen within the first 16 months of observation,
even in the case of a modest network of observatories (e.g., only LIGO-Hanford
and LIGO-Livingston) operating at modest sensitivities (e.g., advanced LIGO
design sensitivity, but without signal recycling mirrors), and assuming a
conservative distribution of beaming angles (e.g. all GRBs beamed with
$\theta_j=30^\circ$). Less conservative assumptions reduce the waiting time
until first detection to a period of weeks to months. Alternatively, the compact binary coalescence
model of short GRBs can be ruled out if a binary is not seen within the first two years of
operation of a LIGO-Hanford, LIGO-Livingston, and Virgo network at
advanced design sensitivity. We also demonstrate that the rate of GRB triggered
sources is less than the rate of untriggered events if
$\theta_j\lesssim30^\circ$, independent of the noise curve, network
configuration, and observed GRB rate. Thus the first detection in GWs of a binary
GRB progenitor is unlikely to be associated with the observation of a GRB.
\end{abstract}
%\keywords{cosmology: theory---galaxies: clusters: general}

%\pacs{95.35.+d,95.36.+x,98.80.Es}% PACS, the Physics and Astronomy
                             % Classification Scheme.
%\keywords{Suggested keywords}%Use showkeys class option if keyword
                              %display desired
\maketitle

\section{Introduction}
\label{sec:intro}

The LIGO and Virgo collaborations have recently released results from roughly a half year of
observations, investigating the gravitational wave (GW) sky at unprecedented levels
of sensitivity~\citep{LIGO:2012aa}. They did not identify any gravitational wave sources, and
thereby established new upper limits on the rates of a variety of possible GW events in
the nearby ($<200\,\mbox{Mpc}$) Universe~\citep{abadie:2011wc}. One of
the most promising sources for GWs detectable by these ground-based observatories
is the coalescence and merger of a compact binary system: two neutron stars (NS), two
black holes (BH), or one of each. 

There has been an active program of observing gamma-ray bursts (GRBs), focusing
on rapid follow-up to determine afterglows and identify host
galaxies~\citep{2006ApJ...650..261S,2006MNRAS.367L..42P,2007ApJ...664.1000B,2009ApJ...696.1871P}. As
a result, there is growing evidence that short/hard gamma-ray bursts are
associated with the mergers of either two neutron stars, or a neutron star with
a black hole~\citep{2010ApJ...708....9F,2011MNRAS.413.2004C,Berger:2011il}. This
consensus is based on noting that the physical timescales are commensurate, the short GRBs
do not appear to be associated with star formation (and therefore are unlikely
to be associated with supernovae), and the GRBs occur far from the centers of their
host galaxies. These studies have also provided redshifts for a subsample of
short GRBs, thereby providing preliminary estimates for the rate densities of these
events~\citep{Nakar:2005bs,Dietz:2011by}. There is great interest in
gravitational wave/electromagnetic multi-messenger observations of these
GRBs \citep{2012ApJ...746...48M,Evans:2012ta,Briggs:2012vj}, as such systems
would help confirm the first detections of GWs, 
elucidate the properties of GRBs, and potentially provide interesting
measurements of the Hubble constant and the dark energy equation-of-state
\citep{Schutz:1986bz,2005ApJ...629...15H,PhysRevD.74.063006,2010ApJ...725..496N}.

One of the most important properties of GRBs is the beaming of the gamma
rays. This beaming directly relates to the total electromagnetic energy of the
explosion, as well as the intrinsic event rate of the sources (as compared to
the observed rate, which is a function of the ones that happen to point at us).
Recent observations of a jet break in the short-duration gamma-ray
burst GRB 111020A suggests a beaming opening angle of
$\theta_j\sim3$--$8^{\circ}$~\citep{Fong:2012wz}. Other GRBs (e.g., GRB 070714B, GRB 070724A, and GRB
071227) have been found with
beaming angles in the range
$1$--$30^{\circ}$~\citep{Fong:2012wz,Coward:2012uz}, while non-detection of a jet
break in the light curve of GRB 050724A places a lower limit on the beaming of
that burst of $\theta_j\ge 25^{\circ}$~\citep{Grupe:2006uc}. Numerical studies, on the other hand, find
$\theta_j\le 30^{\circ}$~\citep{Popham:1998ab,Rosswog:2002rt,Janka:2005yh,Rezzolla:2011da}.

In this paper we estimate the limits that arise on the beaming of short-duration
GRBs based on the non-detection of GWs from associated binary systems in the
recent LIGO/Virgo science run. We also make projections for the detection rate
of binary systems, as a function of mass and beaming angle, for future networks
of GW observatories.
%We know that short GRBs occur, and can estimate their observed rates. The
%intrinsic rate of the progenitor mergers is related to the observed rate through
%a beaming angle; the gamma-rays are beamed, while the gravitational waves are
%not.
We take a conservative lower limit on the observed rate density of local short
GRBs of ${\cal R}_{\rm GRB}= 10\,\mbox{yr}^{-1}\mbox{Gpc}^{-3}$
\citep{Nakar:2005bs,Dietz:2011by,Coward:2012uz}, based primarily on BATSE and
  {\em Swift}\/ observations. We emphasize that this rate is
determined purely through observations, although it is broadly consistent with
the rates arising from population
synthesis~\citep{2006ApJ...648.1110B,OShaughnessy:2008bb,2010ApJ...715L.138B,Abadie:2010fn,Dominik:2012vs}.  The
gravitational-wave limits presented here are based on observed GRB
rates, and are therefore independent of, and complementary to, estimates
based on population synthesis modeling.

We assume that all short GRBs are associated
with low-mass compact binary coalescence. There is developing evidence that
this is the case, with perhaps a small sample of nearby GRBs occurring from
other mechanisms, such as flares from soft gamma
repeaters~\citep{Levan:2008hl,Abbott:2008kg}.
While it is conceivable that not all short GRBs are the result of binary
coalescences, it is perhaps even more likely that not all
binary coalescences result in GRBs. We thus expect that our limits
on the minimum beaming angle in Sec.~\ref{sec:S6} are low, and our estimates of
the maximum wait time in Sec.~\ref{sec:aLIGO} are high.

\section{LIGO S6/Virgo VSR2}
\label{sec:S6}

From July 2009 to October 2010 the LIGO and Virgo observatories conducted a 
search (S6/VSR2--3) for compact binary
coalescences~\citep{abadie:2011wc,abadie:2012jb,LIGO:2012aa}. They did not
detect any 
gravitational-wave events, and thereby established upper limits on the event
rates of coalescences in the local Universe~\citep{abadie:2011wc}. The LIGO instruments operating
during S6 were the $4\,$km laser interferometers at Hanford, WA [H] and
Livingston, LA [L], while the Virgo [V] results were from a single $3\,$km laser
interferometer in Cascina, Italy.
The S6/VSR2 runs consisted of 
0.09 years of HLV coincident data, 0.17 years of HL, 0.10 years of HV, and 0.07
years of LV.\footnote{Due to compromised sensitivity from the
installation of an incorrect mirror, we follow~\citet{abadie:2011wc} in removing
the Virgo SR3 data from our analysis.}

We have taken the representative sensitivities presented in Fig.~1
of~\citet{abadie:2011wc}, and calculated the corresponding horizon distances,
$R_0$, for H, L, and V, where horizon distance is defined to be the distance at
which a given signal-to-noise (SNR), $\rho$, is measured for an optimally oriented
(face-on) and optimally located (directly overhead)
binary. From~\citet{PhysRevD.74.063006}:
\begin{equation} \label{eq:R0}
R_0=4{\cal A}\sqrt{I_{7}}/\rho,
\end{equation}
where ${\cal A}=\sqrt{5/96}\pi^{-2/3}({G{\cal M}/c^3})^{5/6}c$, and where the
binary chirp mass is given by
${\cal M} =(m_1m_2)^{3/5}/(m_1+m_2)^{1/5}$. In this paper we are only interested in nearby
sources ($z<0.2$), and for simplicity neglect the redshift dependence of the chirp
mass~\citep{2005ApJ...629...15H}.
%$\rho_{\rm min}$ is the signal-to-noise
%(SNR) threshold for the detector.
The characteristics of the detector are encapsulated in
\begin{equation}
I_7=\int_{f_{\rm low}}^{f_{\rm high}}\!\frac{f^{-7/3}}{S_h(f)}\,df ,
\end{equation}
with $S_h(f)$ the noise spectral density of the detector.
%The lower limit on frequency
%depends on detectors' ability.
We follow the approach of LIGO's compact binary coalescence
searches~\citep{LIGO:2012aa}, and take
$f_{\rm low}=40\,{\rm Hz}$ for the LIGO detectors, and $f_{\rm low}=50\,{\rm
  Hz}$ for Virgo, with upper limits set by the frequency of the innermost stable
circular orbit.
To calculate the waveform we assume the members of the binary are non-spinning,
and make use of the stationary phase approximation \citep{PhysRevD.74.063006,LIGO:2012aa}.
%We consider a signal-to-noise (SNR) threshold of 8 \daniel{HY: more detail on why
%we choose this and how it's defined? some paper indicates this should correspond
%to a reasonable false alarm rate?} 
%We find that a signal-to-noise (SNR) threshold of $\rho=9.4$ matches the distance horizon plots in Fig.~2
%of~\citet{abadie:2011wc} at low mass.

\begin{figure}
\centering 
\includegraphics[width=0.98\columnwidth]{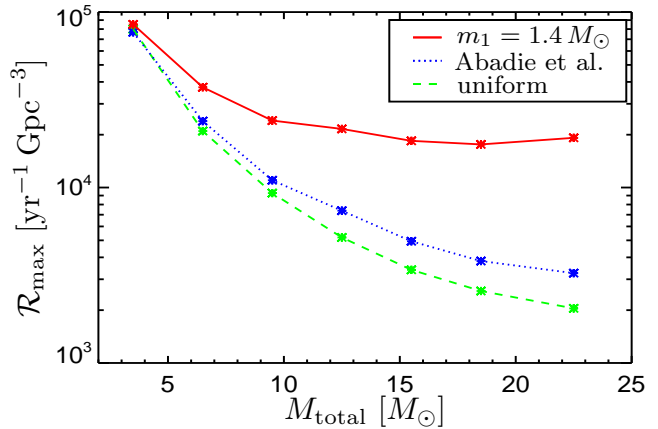}
\caption{\label{fig:rate_S6}
Maximum rate density of binary coalescences as a function of total mass of the
binary, given the non-detection of any
binary systems in the LIGO/Virgo S6/V2 observing runs.
The dotted blue curve is the result from Fig.~4 of \citet{abadie:2011wc}, setting
an upper limit on the rate density, where each mass bin is averaged over a
uniform distribution of component masses. The
dashed green curve shows our result for this curve, where we have assumed a
fixed SNR network threshold of 
$\rho=9.4$. The lower sensitivity of the Abadie curve at higher mass is due to
non-stationary noise. We calculate an effective SNR network threshold, as a
function of mass, to match the Abadie curve, and we use this to calculate the
equivalent curve assume that $m_1=1.4\,M_\odot$ (i.e., that the NS mass is
fixed, and the black hole mass is given by $m_2=M_{\rm
  total}-1.4$).
}
\end{figure}

We follow the approach of~\citet{Schutz:2011fn} to
combine the antenna patterns of the three different interferometers
(HL, HV, LV, and HLV), taking into account the
differing horizon distances (which are a function of the masses of the binary) 
as well as power patterns. The network weighted antenna power pattern,
from eqs.~(15)--(20) of \citet{Schutz:2011fn}, is given by:
%\begin{equation}
\begin{multline} \label{eq:powerpattern}
P_N(\theta,\phi)\equiv\\
\displaystyle\sum\limits_{k=1}^N \left(
(F^2_{+,k}(\theta,\phi,\psi)+F^2_{\times,k}(\theta,\phi,\psi)) \rho_{{\rm
    min},k}^2 R_{0,k}^2\right),
\end{multline}
%\end{equation}
where $F^2_{+,k}(\theta,\phi,\psi)$ and $F^2_{\times,k}(\theta,\phi,\psi)$ are
the antenna patterns for the $k$th detector, and $\rho_{{\rm min},k}$ and $R_{0,k}$ are the SNR
threshold and horizon distances of that detector, respectively. The detection distance for a
network is related to the antenna pattern:
$R(\theta,\phi)=\sqrt{P_N(\theta,\phi)}/\rho_{\rm min}$, where $\rho_{\rm min}$
is the network signal-to-noise threshold. The sensitivity also depends upon
the orientation of the binary; integrating over all orientations results in a factor
of 0.29 decrease in the mean detectable
volume~\citep{Sathyaprakash:2009xs,Schutz:2011fn}:
\begin{equation} \label{eq:volume}
\bar{V}={\frac{0.29}{3} \frac{1}{\rho_{\rm min}^3} \displaystyle\int\! P_N^{3/2}(\theta,\phi)\,d\Omega}.
\end{equation}
%Using this expression we are able to calculate the inverse rate density,
%$\mbox{volume}\times\mbox{time}$, for each detector
%network and observational windows. We set a combined network
%threshold of 12. \daniel{check this.}

Since the LIGO/Virgo network did not detect any gravitational wave sources
during its last science run, we can combine all network
configurations and corresponding coincident observational times to estimate a $90\%$
upper limit to the rate density: ${\cal R}=2.3/(\sum_i\bar V_i\times \Delta
t_i)$, where the sum is over the different detector networks configurations,
$\Delta t_i$ is the amount of observational time for network configuration $i$,
and the factor of 2.3 is in accordance with a Poisson process (see the
discussion below eq.~\ref{eq:time}). We plot our
results in Fig.~\ref{fig:rate_S6}, assuming a combined network threshold of
$\rho=9.4$. This is to be compared with Fig.~4 of \citet{abadie:2011wc},
which calculates the same quantity through detailed analysis of the GW
data stream from the LIGO and Virgo instruments~\citep{2004CQGra..21S1775B}. We have tuned our
SNR threshold to agree for low mass binaries, but our results begin to deviate at higher mass ($M_{\rm
  total}\gtrsim10\,\msun$),
  since the signal shifts to lower frequencies and is therefore more sensitive 
to non-stationary noise in the detectors (in part because there are fewer cycles
to integrate against). To get a sense of the importance of this effect, we
calculate the effective network SNR threshold which we would need to apply, as a
function of mass, to match the rate limits which come out of the full
analysis presented in \citet{abadie:2011wc}. From a value of $\rho=9.4$ at
$M_{\rm total}=3.5\,\msun$,
the non-stationary noise
degrades the sensitivity of the instruments at higher masses, leading to $\rho=10.7$ at
$11$--$14\,\msun$, and $\rho=11.1$ at $20$--$25\,\msun$. In what
follows we incorporate this mass dependence into our effective SNR thresholds. We note that 
\citet{abadie:2011wc} assume a uniform distribution of component masses for their
binaries. We also consider the case where the neutron star is restricted to have
$m_1=1.4\,\msun$, and the mass of the 
companion is given by $m_2=M_{\rm total}-m_1$. Because this entails
higher mass ratios for higher mass binaries, it decreases the overall
gravitational-wave strength of the
sources in comparison to the uniform distributions, and therefore decreases
the detectable volume. As can be seen in Fig.~\ref{fig:rate_S6}, this results in
a negligible effect at low mass, but rises to a factor of 3.6 in the rate at
$M_{\rm total}=15\,\msun$ and 6.8 at $M_{\rm total}=25\,\msun$.

\begin{figure}[t]
\centering 
\includegraphics[width=0.98\columnwidth]{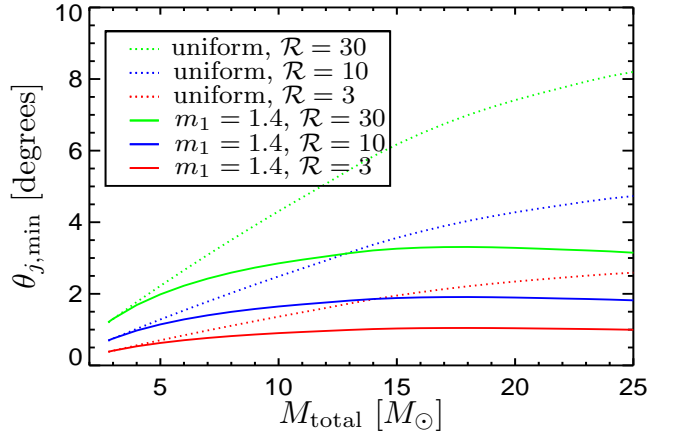}
\caption{\label{fig:beaming_S6}
Minimum beaming half-angle, $\theta_{j,\rm min}$, as a function of the total mass
of the binaries. We plot results for three different rates of short GRBs, ${\cal R}=$ 3, 10,
and $30\,\mbox{yr}^{-1}\mbox{Gpc}^{-3}$, and for two different distributions
($m_1$ is uniform from 1 to $M_{\rm total}$, and $m_1=1.4\,M_\odot$).
Given the lack of detection of binaries in LIGO/Virgo S6/V2,
the beaming angles of short GRBs will be greater than what is plotted in 90\%
of cases. We use a network threshold which matches the \citet{abadie:2011wc}
results (see text for details).
}
\end{figure}

We define the beaming angle, $\theta_j$, to be the half opening
angle of one of the two polar jets of a gamma-ray burst. The fraction of
the sky, $f_b$, covered by the beamed gamma rays is given by $f_b={1-\cos\theta_j}$.
Given the paucity of data on the beaming of short GRBs, it
is premature to assume knowledge of the distribution of beaming angles. We
therefore will assume that all short GRBs have a fixed beaming angle, $\theta_j$, with the
understanding that this fixed angle provides the same results as the appropriate
average of the true distribution of beaming angles. In other words,
$1/(1-\cos\theta_j)\equiv\int\!P(\theta)/(1-\cos\theta)\,d\theta$, where
$P(\theta)$ is the true distribution of beaming angles. 
Assuming all short GRBs have
compact binary progenitors, the implied rate density of these coalescences is
given by ${\cal R} = {\cal R}_{\rm GRB}/f_b={\cal R}_{\rm
  GRB}/(1-\cos\theta_j)$. 
If only a fraction, $f_{\rm CBC}$,
of short GRBs result from compact binary coalescence, then the rate density
becomes ${\cal R} = f_{\rm CBC}{\cal R}_{\rm  GRB}/(1-\cos\theta_j)$.
In Fig.~\ref{fig:beaming_S6} we plot the 90\% lower limit on the beaming angle as a function
of the mass of the progenitors. We take the observed rate of short GRBs to be 
3, 10, or $30\,{\rm yr}^{-1}{\rm Gpc}^{-3}$
\citep{Nakar:2005bs,2009A&A...498..329G,Coward:2012uz}.
For example, a rate of $3\,{\rm yr}^{-1}{\rm Gpc}^{-3}$ can be thought of as a
very conservative estimate of ${\cal R}=2\,{\rm yr}^{-1}{\rm 
  Gpc}^{-3}$ with a very conservative estimate of $f_{\rm CBC}=0.5$.
This simple analysis suggests that models of
short GRBs with progenitors of mass $M_{\rm total}>20\,\msun$ (uniformly distributed in
component mass) and with a beaming angle $\theta_j<4^\circ$ are inconsistent 
with existing LIGO/Virgo data. This weakens significantly for more
realistic masses and mass ratios, with a minimum beaming angle of $\sim1^\circ$
at $M_{\rm total}\sim3\,\msun$. These limits are completely consistent with
observations and expectations. The current LIGO/Virgo
data is on the verge of providing interesting astrophysical constraints,
which suggests that the next generation of detectors should provide quick
detections, or alternatively, the lack of quick detections would provide strong
lower limits on the beaming of short GRBs. We explore these constraints in the next section.

\section{Advanced LIGO/Virgo}
\label{sec:aLIGO}

We now calculate the expected detection rate of short GRB progenitors in the advanced LIGO and Virgo
detectors, as well as additional detectors in Japan (KAGRA)\footnote{{\tt gwcenter.icrr.u-tokyo.ac.jp/en/}} and India
(IndIGO)\footnote{{\tt www.gw-indigo.org/tiki-index.php}}.
The advanced LIGO detectors are expected to begin operation in $\sim2015$, and it is
hoped that the LIGO and Virgo observatories will achieve their target  advanced detector
sensitivities by $\sim2017$, with the Japanese [J] and Indian [I] detectors operating at
comparable sensitivities by $\sim2020$. We assume an identical
noise curve for each of these instruments, given by the representative advanced
LIGO noise curves in LIGO document T0900288-v3,\footnote{{\tt
dcc.ligo.org/cgi-bin/DocDB/ShowDocument?docid=2974}} with $f_{\rm low}=10\,{\rm Hz}$.
 We take the target design
sensitivity to be given by the
{\tt ZERO\_DET\_high\_P.txt} curve, corresponding to zero-detuning of the signal recycling mirror, and
high laser power. We also consider an early, less sensitive incarnation of
the detectors resulting from the absence of signal recycling mirrors, given by the {\tt NO\_SRM.txt}
curve. It is possible that in 2015 the worldwide GW detector
network will consist solely of HL in this lower sensitivity configuration.

We calculate the mean detectable volume, $\bar{V}$, of a variety of ground-based
networks. Our results are presented in Table~\ref{table:networks}, where in all
cases we have assumed a network SNR threshold of $\rho=10$, and our sources are
taken to be equal-mass binaries with $m_1=m_2=1.4\,\msun$.
The mean detectable volume of a network is expected to scale as
${\cal M}^{5/2}$ (see eqs.~(\ref{eq:R0}), (\ref{eq:powerpattern}), and
(\ref{eq:volume})), although this relation is imperfect, since the scaling also depends on the shape of the
noise curve.
%For example, sufficiently massive binaries merge 
%below the seismic noise floor of the detectors, and are undetectable.
We fit the mass dependence to the functional form
\begin{equation}
\bar{V}(M_{\rm total}) = V_0\left({\frac{M_{\rm total}}{2.8\,\msun}}\right)^p,
\label{eq:mass_fit}
\end{equation}
where $V_0$ is the detection volume for a binary with $M_{\rm
total}=2.8\,\msun$, and we follow the previous section and fix $m_1=1.4\,\msun$ and $m_2=M_{\rm total}-m_1$.
The ``no SRM'' noise curve yields $p=1.39$, while our fiducial advanced LIGO
noise curve yields $p=1.30$. These fits are good to 20\% at $M_{\rm
  total}=30\,\msun$.\footnote{For completeness, we also mention results for the
  {\tt ZERO\_DET\_low\_P.txt} noise curve, corresponding to a lower laser
  power. We find $V_0=0.02\,\mbox{Gpc}^3$ for HL (no SRM),
  $V_0=0.034\,\mbox{Gpc}^3$ for HLV, and $p=1.37$.}

\begin{table}[tp]%
\centering
\begin{tabular}{clccccc}
\toprule
        Network     & $V_{0}(\rm{Gpc}^{3})$  	& $T_{\rm first}$	& $\theta_j=10^\circ$	& $\theta_j=30^\circ$	& $\theta_j=90^\circ$	\\ \midrule
        HL(no SRM)  & $0.027$    				& $1.00$			& $1.8$					& $16$					& $120$	\\ 
        HLV(no SRM) & $0.046$			    	& $0.59$			& $1.1$					& $9.4$					& $70$	\\ 
        HLV		   	& $0.092$    				& $0.31$			& $0.56$				& $4.9$					& $37$	\\ 
        HLVJ		& $0.14$    				& $0.21$			& $0.37$				& $3.3$					& $25$	\\ 
        HLVI		& $0.14$    				& $0.20$			& $0.36$				& $3.2$					& $24$	\\ 
        HLVJI		& $0.19$					& $0.15$ 			& $0.27$				& $2.3$					& $17$	\\ \bottomrule
\end{tabular}
\caption{Mean detectable volume and wait times for the detection of binary
  coalescence associated with short GRBs in future GW detector
  networks. The network SNR is taken to be 10, and the volume is calculated for
  a $1.4\,\msun$--$1.4\,\msun$ binary. $T_{\rm first}$ is the waiting time until
  first detection, scaled to the value for the HL network with the ``no SRM''
  noise curve. The last three columns list the $90\%$
  wait time for first detection (in months) for three different values of the
  beaming angle.
\label{table:networks}}
\end{table}

We are interested in how quickly networks of advanced ground-based
gravitational-wave detectors can be expected to see their first binary
coalescences associated with short GRB progenitors. 
This rate is a function of the rate of observed GRBs
(in gamma rays), ${\cal R}_{\rm GRB}$, the sensitivity and configuration of the
detectors, the mass distribution of the GRB progenitors, and the beaming of the
GRBs. The event rate of detectable binaries for a network of GW observatories is given by
\begin{equation}
\lambda=\bar{V}(M_{\rm total}){\cal R}_{\rm GRB} f_{\rm CBC}/(1-\cos\theta_j).
\label{eq:time}
\end{equation}
How long will a given network have to wait before seeing its first event?
This is described by a Poisson process, with the probability of
waiting a time $\tau$ before detecting the first event given by
$e^{-\tau\lambda}$. We define $t_{\rm first}$ as the waiting time by which, in
90\% of cases, the first event will have been observed: $t_{\rm
  first}=-\ln(0.1)/\lambda=2.3/\lambda$.  In Fig.~\ref{fig:t_aLIGO} we plot
$t_{\rm first}$ as a function of the beaming angle, $\theta_j$, for the HL and
HLV networks. If one is interested in the waiting time by which the first event
has been seen in 50\% or 99\% of the cases, the 90\% waiting times are multiplied
by 0.3 or 2, respectively. 
We have assumed $f_{\rm CBC}\times{\cal R}_{\rm GRB}=
10\,\mbox{yr}^{-1}\mbox{Gpc}^{-3}$, and we have considered NS-NS equal mass
binaries, with $m_1=m_2=1.4\,\msun$.
%% We consider two possible noise curves for the detectors: ``no
%% SRM'' and ``HIGH''. NO SRM corresponds to a configuration without 
%% a signal recycling mirror, and might represent an early, less sensitive first
%% incarnation of the Advanced LIGO detectors. The LOW and HIGH configurations
%% represent a configuration with zero-detuning of the signal recycling mirror, and
%% low and high laser power, respectively.
We do not employ the mass-dependent threshold correction factors from the
S6/V2--3 analysis derived in the previous section, since it can be argued that the
non-stationary noise is less likely to
be a problem in the advanced configuration of these detectors,\footnote{Peter
Shawhan, private communication.} and, regardless, the functional form would
not {\it a priori}\/ be expected to match that of the lower sensitivity
detectors.
The representative curves in Fig.~\ref{fig:t_aLIGO} can be rescaled to
other parameter values of interest.
For example, if either ${\cal R}_{\rm GRB}$ or $f_{\rm CBC}$ is down by a factor of
10, then the waiting times, $t_{\rm first}$, are all multiplied
by the same factor of 10.
A change in the network configuration similarly results in an overall
shift in the expected rates, and therefore an overall shift in $t_{\rm first}$.
The relative waiting times for other networks, $T_{\rm first}$, are presented in
Table~\ref{table:networks}, scaled to the HL (no SRM) value.
The HLVJI network, with all detectors at the advanced zero-detuning high laser power sensitivity,
has a waiting time that is a factor of 0.15 that of the HL (no SRM) curve, independent of
the beaming angle.
If we consider larger mass systems, the waiting time is correspondingly
shortened (see eqs.~(\ref{eq:mass_fit}) and~(\ref{eq:time})).
For example, if we consider NS--BH systems, with $M_{\rm NS}=1.4\msun$
and with $M_{\rm BH}=10\,\msun$ and $20\,\msun$, then the waiting times until
first detection are a factor of 0.11 and 0.05 shorter than those for the NS--NS case
presented in Fig.~\ref{fig:t_aLIGO}. It is to be noted that our results are
roughly consistent with the completely independent rate estimates from
population synthesis and observed binary pulsars~\citep{Abadie:2010fn}.

Alternatively, the binary origin of short GRBs can be falsified (at the 90\%
Poisson confidence discussed above) if no
coalescences are
observed with a full network (HLV)
operating at design sensitivity (zero-detuning high laser power) over a period
of 3 years. This limit comes from taking $\theta_j=90^\circ$, which
corresponds to none of the GRBs being beamed (which is already inconsistent with
observations).
If we take a conservative upper limit of $\theta_j=45^\circ$, we
find that the binary origin can be falsified (at the 99\% level) in 70 months for HL (no SRM), and
in 22 months for HLV.
However, even if short GRBs are not the result of binary mergers,
we nonetheless expect a population of merging systems, and
these should be observable by future observatories~\citep{Abadie:2010fn}.

The predicted event rates for the various networks are related to the 90\%
waiting times by a factor of 2.3. If an HL network, using conservative noise
curves (no SRM), takes 16 months before a first detection assuming a
binary of mass $1.4\,\msun$--$1.4\,\msun$ and a beaming angle of $\theta_j=30^\circ$ (see
Table~\ref{table:networks}), then the predicted event rate for this network is
$2.3/1.33\,\mbox{yr}=1.7\,\mbox{yr}^{-1}$. These results are consistent with
those of \citet{Coward:2012uz}.

%\medskip

\section{Triggered versus Untriggered}
\label{sec:trigger}

\begin{figure}
\centering 
\includegraphics[width=0.98\columnwidth]{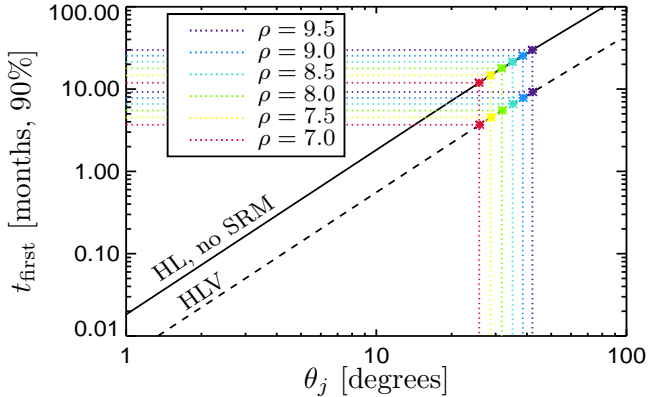}
\caption{\label{fig:trigger}
\label{fig:t_aLIGO}
Waiting time until first detection, $t_{\rm first}$, as a function of the
beaming angle, $\theta_j$. Results are shown for both untriggered and triggered
GW observations of GRB progenitors, where we have assumed the observed local
short GRB rate is 
${\cal R}_{\rm GRB}= 10\,\mbox{yr}^{-1}\mbox{Gpc}^{-3}$.
In 90\% of cases the waiting time will be less than
the values indicated. We plot results for two different GW networks:
Hanford+Livingston, operating without a signal recycling mirror (a potential
early version of advanced LIGO) and Hanford+Livingston+Virgo, operating at the
design sensitivity (at high laser power). Even in the pessimistic
case (HL, no SRM, all short GRBs are beamed with $\theta_j=30^\circ$), the first
untriggered binary detection is expected in less than 16 months. For more
substantial GW networks, the expected wait time may be less than a month (e.g.,
if $\theta_j\sim10^\circ$).
The dotted lines show the wait times for GRB triggered GW observations, for a
range of values of $\rho$.  The GRB trigger provides a time and sky position,
thereby reducing the required SNR threshold and increasing the detection rate
because the sources are assumed to be face-on. On the other hand, the triggering
GRB rate is given by the $\theta_j=90^\circ$ values, and is not enhanced by
beaming. The dotted lines show the equivalence beaming angles, at which the
rates of triggered and untriggered GW observations of GRBs match each other;
this is found to happen at $\theta_j\sim30^\circ$ for $\rho\sim7.5$. For lower
values of $\theta_j$, untriggered observations occur more frequently than those
triggered by GRBs, and the first observed binary GRB progenitor will be seen
first (and perhaps only) in GWs.  }
\end{figure}

An important question when considering the future GW detection of short GRBs is
whether triggered or untriggered detections are more likely. A short GRB trigger
improves the
sensitivity of the GW search by reducing the need to marginalize
over all times and sky positions. In addition, because the gamma-rays are
thought to be beamed, a GRB trigger is expected to be face-on, thereby
increasing the signal in GWs over a source with a random inclination. In other
words, the strongest GW emitters happen to also be the ones observable in
gamma-rays. A GW network is therefore substantially more sensitive to a GRB triggered source.
On the other hand, for small beaming
angles the rate density of GRB progenitors increases (approaching $\infty$ as
$\theta_j\to0$). For sufficiently small $\theta_j$ the untriggered GW
detection of a GRB progenitor will dominate over a triggered GRB, even
considering the additional sensitivity in the latter case. We are interested in
establishing whether triggered or untriggered GRBs are more likely for upcoming networks
of GW observatories.

We have calculated the waiting time and event rates for untriggered observations
of GRB progenitor systems above. We now consider the equivalent calculation in
the case of a GRB trigger. We assume that the GRB is face-on, which improves the
sensitivity of the network by a factor of $1/0.29$ (see
eq.~(\ref{eq:volume})). We now estimate the reduction in SNR threshold due to
the known time and sky position of the source~\citep{PhysRevD.74.063006}. We assume
$\exp({-\rho^2/2})\propto1/\mbox{\# of templates}$, where in the
untriggered case we took $\rho=10$.
%If the corresponding false alarm rate was set to $10^{-4}$,
%this corresponds to $\sim10^18$ templates.
If we take this threshold to have been based on roughly one year of observation,
the existence of a GRB trigger now reduces the observational window down to
$\sim10\,\mbox{sec}$, for a reduction in the number of templates by a factor of
$\sim10^6$. If the sky localization in the untriggered case is
$\sim5\,\mbox{deg}$~\citep{2011ApJ...739...99N}, then compared to a full-sky
search ($41,253\,\mbox{deg}^2$), the reduction in number of templates is a
factor of $\sim10^3$. The total number of templates is down by a factor of
$10^{9}$, which for the equivalent false alarm rate would imply that the SNR
threshold is reduced to $\rho\sim\sqrt(-2\log(\exp(-10^2/2)\times10^{9}))=7.7$.

In Fig.~\ref{fig:trigger} we compare the wait times for untriggered and
triggered GRBs. We find that the wait time for a triggered GRB is equivalent to
the untriggered case if the average GRB beaming angle is
$\theta_j\sim30^\circ$. This result is independent of the network configuration,
the individual noise curves, and the assumed assumed GRB rate, and is weakly
dependent on the specific value of the 
threshold improvement due to the reduced number of templates. If the GRBs have an
average beaming value of $\theta_j=20^\circ$, we would predict the rate of untriggered
GRBs to be roughly double that of triggered ones. This increases to a factor of
10 if $\theta_j=10^\circ$. Alternatively, the rates are equal if
$\theta_j=30^\circ$, and triggered GRBs are found at double the rate of
untriggered ones if $\theta_j=45^\circ$. It is to be noted that this process can
be inverted, and the wait time
before first detection (and in between the first few detections) may be used to
infer the beaming angle of GRBs. In addition, the relative rates of triggered
and untriggered GRBs will help establish the beaming, and will be an important
test of consistency when compared with explicit determinations of the beaming
distribution based on GW measurements of the inclination of GRB sources.

As discussed in Sec.~\ref{sec:intro}, recent
observations have measured short GRBs with $\theta_j\sim10^\circ$, indicating that
it is likely that the rate of untriggered GRBs will be greater than the
rate of triggered ones, and implying that the first detection of a binary
system which is a progenitor of a short GRB will not be triggered by a
GRB. Although triggered GRBs may be less frequent than untriggered ones,
multi-messenger observations of these systems holds tremendous scientific
potential, and should be aggressively
pursued~\citep{2009arXiv0902.1527B}. Furthermore, the increase in
psychological confidence of detection given coincident observation may play a
large role in the initial detections. It is to be
emphasized that our triggered rates assume the existence of an all-sky short GRB
monitor operating contemporaneously with advanced GW networks.

\section{Discussion}

We have explored the connection between the
observed short GRB rate, the beaming angle of short GRBs, and the predicted rate
of detectable binary systems associated with progenitors of GRBs in networks of
gravitational-wave observatories.

We have shown that existing LIGO/Virgo data provides preliminary constraints on
the beaming angle and mass distribution of short GRB progenitor systems. 
For example, we find that short GRB
progenitors of mass $M_{\rm total}>20\,\msun$ (uniformly distributed in
component mass) and with beaming angles of $\theta_j<4^\circ$ are ruled out
by existing LIGO/Virgo data. These constraints, while novel, are fully
consistent with our current understanding of the short GRB engine and rates.

We have analyzed the observed rate of short GRB progenitors in future networks
of GW detectors. We find that,
even in the pessimistic case of only two detectors (HL) operating at
conservative sensitivity (without a signal recycling mirror), in 90\% of
cases we would expect
a first detection of a binary within 16 months if the GRBs are
beamed within $\theta_j=30^\circ$, and within 55 days if
$\theta_j=10^\circ$. The expected event rates are $1.7\,\mbox{yr}^{-1}$
($\theta_j=30^\circ$) and $15\,\mbox{yr}^{-1}$ 
($\theta_j=10^\circ$). We find that the HLV network, operating at zero-detuning
and high laser power, would shorten these times to 4.9 months ($\theta_j=30^\circ$) and
17 days ($\theta_j=10^\circ$), with corresponding event rates of
$5.6\,\mbox{yr}^{-1}$ and $49\,\mbox{yr}^{-1}$. Alternatively, the
  binary coalescence model for short GRB progenitors can be ruled out if an HLV
  network does not observe a binary within the first two years of observation.
%% We emphasize that our calculations require knowledge of the observed short GRB
%% rate. We have assumed a rate of ${\cal R}_{\rm GRB}=
%% 10\,\mbox{yr}^{-1}\mbox{Gpc}^{-3}$, with 100\% of short GRBs resulting from
%% bianry coalescence. If, for example, the value of ${\cal R}_{\rm GRB}\times
%% f_{\rm CBC}$ is reduced by a factor of 10, then the computed rates are reduced
%% by a factor of 10, and the waiting times are all increased by a factor of 10.

Finally, we have shown that the rate of GRB triggered observations of GW systems
associated with GRBs is lower than the rate of untriggered observations if
$\theta_j\gtrsim30^\circ$. This result is independent of network, noise
curve, and GRB rate, and when coupled with recent observations of small beaming
angles for short GRBs, suggests that the first detections of GRB progenitors with
advanced GW networks will not involve the observation of GRBs.

We conclude that, assuming short GRBs are the result of the merger of compact
objects, and assuming that the resulting gamma-rays are beamed, the first detection of
gravitational-waves from binary coalescence associated with a GRB progenitor
will be untriggered, and may occur within months of operation
of a modest network of ground-based gravitational wave observatories.

\begin{acknowledgments}
%\bigskip
%\medskip
We acknowledge valuable discussions with Edo Berger, Laura Cadonati, Curt
Cutler, Wen-fai Fong, Peter Shawhan, and Rai Weiss.
\end{acknowledgments}

%\appendix

%\section{Appendixes}

\bibliography{references}% Produces the bibliography via BibTeX.

\begin{thebibliography}{39}
\expandafter\ifx\csname natexlab\endcsname\relax\def\natexlab#1{#1}\fi

\bibitem[{Abadie {et~al.}(2010)}]{Abadie:2010fn}
Abadie, J., {et~al.} 2010, Class. Quantum Grav., 27, 173001

\bibitem[{Abadie {et~al.}(2011)}]{abadie:2011wc}
---. 2011, arXiv:1111.7314

\bibitem[{Abadie {et~al.}(2012{\natexlab{a}})}]{abadie:2012jb}
---. 2012{\natexlab{a}}, \aap, 539, A124

\bibitem[{Abadie {et~al.}(2012{\natexlab{b}})}]{LIGO:2012aa}
---. 2012{\natexlab{b}}, arXiv:1203.2674

\bibitem[{Abbott {et~al.}(2008)}]{Abbott:2008kg}
Abbott, B., {et~al.} 2008, \apj, 681, 1419

\bibitem[{Belczynski {et~al.}(2010)Belczynski, Dominik, Bulik, O'Shaughnessy,
  Fryer, \& Holz}]{2010ApJ...715L.138B}
Belczynski, K., Dominik, M., Bulik, T., {et~al.} 2010, \apjl, 715, L138

\bibitem[{{Belczynski} {et~al.}(2006){Belczynski}, {Perna}, {Bulik},
  {Kalogera}, {Ivanova}, \& {Lamb}}]{2006ApJ...648.1110B}
{Belczynski}, K., {Perna}, R., {Bulik}, T., {et~al.} 2006, \apj, 648, 1110

\bibitem[{Berger(2011)}]{Berger:2011il}
Berger, E. 2011, New A Rev., 55, 1

\bibitem[{Berger {et~al.}(2007)Berger, Fox, Price, Nakar, Gal-Yam, Holz,
  Schmidt, Cucchiara, Cenko, Kulkarni, Soderberg, Frail, Penprase, Rau, Ofek,
  Burnell, Cameron, Cowie, Dopita, Hook, Peterson, Podsiadlowski, Roth,
  Rutledge, Sheppard, \& Songaila}]{2007ApJ...664.1000B}
Berger, E., Fox, D.~B., Price, P.~A., {et~al.} 2007, \apj, 664, 1000

\bibitem[{Bloom {et~al.}(2009)Bloom, Holz, Hughes, Menou, Adams, Anderson,
  Becker, Bower, Brandt, Cobb, Cook, Corsi, Covino, Fox, Fruchter, Fryer,
  Grindlay, Hartmann, Haiman, Kocsis, Jones, Loeb, Marka, Metzger, Nakar,
  Nissanke, Perley, Piran, Poznanski, Prince, Schnittman, Soderberg, Strauss,
  Shawhan, Shoemaker, Sievers, Stubbs, Tagliaferri, Ubertini, \&
  Wozniak}]{2009arXiv0902.1527B}
Bloom, J.~S., Holz, D.~E., Hughes, S.~A., {et~al.} 2009, arXiv:0902.1527

\bibitem[{{Brady} {et~al.}(2004){Brady}, {Creighton}, \&
  {Wiseman}}]{2004CQGra..21S1775B}
{Brady}, P.~R., {Creighton}, J.~D.~E., \& {Wiseman}, A.~G. 2004, Classical and
  Quantum Gravity, 21, 1775

\bibitem[{Briggs {et~al.}(2012)}]{Briggs:2012vj}
Briggs, M.~S., {et~al.} 2012, arXiv:1205.2216

\bibitem[{Church {et~al.}(2011)Church, Levan, Davies, \&
  Tanvir}]{2011MNRAS.413.2004C}
Church, R.~P., Levan, A.~J., Davies, M.~B., \& Tanvir, N. 2011, \mnras, 413,
  2004

\bibitem[{Coward {et~al.}(2012)Coward, Howell, Piran, Stratta, Branchesi,
  Bromberg, Gendre, Burman, \& Guetta}]{Coward:2012uz}
Coward, D., Howell, E., Piran, T., {et~al.} 2012, arXiv:1202.2179

\bibitem[{Dalal {et~al.}(2006)Dalal, Holz, Hughes, \&
  Jain}]{PhysRevD.74.063006}
Dalal, N., Holz, D.~E., Hughes, S.~A., \& Jain, B. 2006, \prd, 74, 063006

\bibitem[{Dietz(2011)}]{Dietz:2011by}
Dietz, A. 2011, \aap, 529, A97

\bibitem[{Dominik {et~al.}(2012)Dominik, Belczynski, Fryer, Holz, Berti, Bulik,
  Mandel, \& O'Shaughnessy}]{Dominik:2012vs}
Dominik, M., Belczynski, K., Fryer, C., {et~al.} 2012, arXiv:1202.4901

\bibitem[{Evans {et~al.}(2012)}]{Evans:2012ta}
Evans, P.~A., {et~al.} 2012, arXiv:1205.1124

\bibitem[{Fong {et~al.}(2010)Fong, Berger, \& Fox}]{2010ApJ...708....9F}
Fong, W., Berger, E., \& Fox, D.~B. 2010, \apj, 708, 9

\bibitem[{Fong {et~al.}(2012)Fong, Berger, Margutti, Zauderer, Troja, Czekala,
  Chornock, Gehrels, Sakamoto, Fox, \& Podsiadlowski}]{Fong:2012wz}
Fong, W., Berger, E., Margutti, R., {et~al.} 2012, arXiv:1204.5475

\bibitem[{Grupe {et~al.}(2006)Grupe, Burrows, Patel, Kouveliotou, Zhang,
  {et~al.}}]{Grupe:2006uc}
Grupe, D., Burrows, D.~N., Patel, S.~K., {et~al.} 2006, \apj, 653, 462

\bibitem[{{Guetta} \& {Stella}(2009)}]{2009A&A...498..329G}
{Guetta}, D., \& {Stella}, L. 2009, \aap, 498, 329

\bibitem[{Holz \& Hughes(2005)}]{2005ApJ...629...15H}
Holz, D.~E., \& Hughes, S.~A. 2005, \apj, 629, 15

\bibitem[{Janka {et~al.}(2006)Janka, Mazzali, Aloy, \& Pian}]{Janka:2005yh}
Janka, H.-T., Mazzali, P., Aloy, M.-A., \& Pian, E. 2006, \apj, 645, 1305

\bibitem[{Levan {et~al.}(2008)Levan, Tanvir, Jakobsson, Chapman, Hjorth,
  Priddey, Fynbo, Hurley, Jensen, Johnson, Gorosabel, Castro-Tirado, Jarvis,
  Watson, \& Wiersema}]{Levan:2008hl}
Levan, A.~J., Tanvir, N.~R., Jakobsson, P., {et~al.} 2008, \mnras, 384, 541

\bibitem[{{Metzger} \& {Berger}(2012)}]{2012ApJ...746...48M}
{Metzger}, B.~D., \& {Berger}, E. 2012, \apj, 746, 48

\bibitem[{Nakar {et~al.}(2006)Nakar, Gal-Yam, \& Fox}]{Nakar:2005bs}
Nakar, E., Gal-Yam, A., \& Fox, D.~B. 2006, \apj, 650, 281

\bibitem[{Nissanke {et~al.}(2010)Nissanke, Holz, Hughes, Dalal, \&
  Sievers}]{2010ApJ...725..496N}
Nissanke, S., Holz, D.~E., Hughes, S.~A., Dalal, N., \& Sievers, J.~L. 2010,
  \apj, 725, 496

\bibitem[{{Nissanke} {et~al.}(2011){Nissanke}, {Sievers}, {Dalal}, \&
  {Holz}}]{2011ApJ...739...99N}
{Nissanke}, S., {Sievers}, J., {Dalal}, N., \& {Holz}, D. 2011, \apj, 739, 99

\bibitem[{O'Shaughnessy {et~al.}(2008)O'Shaughnessy, Kim, Kalogera, \&
  Belczynski}]{OShaughnessy:2008bb}
O'Shaughnessy, R., Kim, C., Kalogera, V., \& Belczynski, K. 2008, \apj, 672,
  479

\bibitem[{Panaitescu(2006)}]{2006MNRAS.367L..42P}
Panaitescu, A. 2006, \mnras: Letters, 367, L42

\bibitem[{Perley {et~al.}(2009)Perley, Metzger, Granot, Butler, Sakamoto,
  Ramirez-Ruiz, Levan, Bloom, Miller, Bunker, Chen, Filippenko, Gehrels,
  Glazebrook, Hall, Hurley, Kocevski, Li, Lopez, Norris, Piro, Poznanski,
  Prochaska, Quataert, \& Tanvir}]{2009ApJ...696.1871P}
Perley, D.~A., Metzger, B.~D., Granot, J., {et~al.} 2009, \apj, 696, 1871

\bibitem[{Popham {et~al.}(1999)Popham, Woosley, \& Fryer}]{Popham:1998ab}
Popham, R., Woosley, S., \& Fryer, C. 1999, \apj, 518, 356

\bibitem[{Rezzolla {et~al.}(2011)Rezzolla, Giacomazzo, Baiotti, Granot,
  Kouveliotou, {et~al.}}]{Rezzolla:2011da}
Rezzolla, L., Giacomazzo, B., Baiotti, L., {et~al.} 2011, \apj, 732, L6

\bibitem[{Rosswog \& Ramirez-Ruiz(2002)}]{Rosswog:2002rt}
Rosswog, S., \& Ramirez-Ruiz, E. 2002, \apj, 336, L7

\bibitem[{Sathyaprakash \& Schutz(2009)}]{Sathyaprakash:2009xs}
Sathyaprakash, B., \& Schutz, B. 2009, Living Rev.Rel., 12, 2

\bibitem[{Schutz(1986)}]{Schutz:1986bz}
Schutz, B.~F. 1986, \nat, 323, 310

\bibitem[{Schutz(2011)}]{Schutz:2011fn}
---. 2011, Class. Quantum Grav., 28, 125023

\bibitem[{Soderberg {et~al.}(2006)Soderberg, Berger, Kasliwal, Frail, Price,
  Schmidt, Kulkarni, Fox, Cenko, Gal-Yam, Nakar, \& Roth}]{2006ApJ...650..261S}
Soderberg, A.~M., Berger, E., Kasliwal, M., {et~al.} 2006, \apj, 650, 261

\end{thebibliography}

\end{document}